\documentclass[a4paper]{jpconf}
\usepackage{graphicx}
\begin{document}
\title{The effective interaction hyperspherical harmonics method for non-local potentials}

\author{G. Orlandini$^{1,2}$, N. Barnea$^{3}$ and W. Leidemann$^{1,2}$}

\address{$^1$ Dipartimento di Fisica, Universit\`a di Trento, Via Sommarive 14, I-38123 Trento, Italy\\
$^2$ INFN, Gruppo Collegato di Trento, I-38123 Trento, Italy, \\
$^3$ The Racah Institute of Physics, The Hebrew University, 91904 Jerusalem, Israel
}

\ead{orlandin@science.unitn.it}

\begin{abstract}
A different formulation of the effective interaction hyperspherical harmonics (EIHH) method, suitable 
for  non-local potentials, is presented. The EIHH method for local interactions is first shortly reviewed 
to point out the problems
of an extension to non-local potentials. A viable solution is proposed and, as an application, 
results on the ground-state properties of 4- and 6-nucleon systems are presented. 
One finds a substantial acceleration 
in the convergence rate of the hyperspherical harmonics series. 
Perspectives for an application to scattering cross sections, via the Lorentz 
transform method are discussed. 
\end{abstract}

\section{Introduction}

In solving the Schr\"odinger equation for the ground state of a many body system,
via expansion techniques, one very often encounters the problem of convergence.
In order to deal with this problem already several decades ago the notion of {\it effective interaction} (EI) 
has been introduced. The true EI is the operator that replaces the bare interaction when working in a finite model 
space, to give (at least) the same ground-state energy. In general it is an $A$-body operator. Therefore in the following 
we will denote it by $V^{[A]eff}$ and refer to it as to the {\it total} effective interaction.
Finding $V^{[A]eff}$, however, is as difficult as solving the original problem, 
therefore over the years one has searched - more or less successfully - for the best approximations to it. 

A different point of view has been introduced in the nineties with the No Core Shell Model (NCSM)~\cite{Navratil96a}.
The basic philosophy behind the NCSM success is that the EI is  a {\it partial} effective interaction, 
tailored for the model space, however, constrained to coincide with the bare one, 
when enlarging the model space. This ensures that the 
energy will converge faster to the correct value.  
So, instead of pointing at getting the best approximation 
to the binding energy at a fixed model space, one points at 
reaching the converged solution as fast as possible, increasing the model space.

The same idea is behind the so called EIHH method~\cite{EIHH}. 
The difference is that, instead of the harmonic oscillator basis (HO), the hyperspherical harmonics (HH) basis is used. 
There are some advantages in the  EIHH approach: the EI is a function of the hyperradius parameter 
(a kind of ``hyperlocal`` effective interaction) and in addition it is state dependent, 
resulting in a faster convergence, in the relevant quantum numbers, than in NCSM. 
The disadvantage is that working with the HH basis one does not take advantage of the well known versatility of the HO basis.
In particular the straightforward application of the EIHH method to non-local potential presents some problems. 
In the following it will be shown what these problems are and how they can be overcome.

\section{Short summary of the EIHH method for local potentials}
\subsection{Hyperspherical coordinates and hyperspherical harmonics}
One characteristic of all the HH methods is that they respect translational invariance.
In fact the {\it hyper}-spherical coordinates are defined by a transformation 
of the Jacobi coordinates ${\vec \eta}_i\,\,\, (i=1,2,...\,, A-1)$, 
in analogy with the spherical coordinates $(r,\hat\Omega_2)$ of the 2-particle case. 

Of the $(3A-3)$ coordinates the transformation retains the $(2A-2)$ Jacobi vector angles. The remaining $(A-1)$
coordinates are obtained  generalizing the 3-dimensional space into a  
$(A-1)$-dimensional space, spanned by the modulus of the $(A-1)$ Jacobi vectors. Therefore,  
instead of the square radius $r^2$, 
that is the sum of the three squared coordinates, 
one has a square  {\it hyper}-radius, $\rho^2$ that is the sum of the square modulus of the $(A-1)$
Jacobi vectors. 
The remaining $(A-2)$ coordinates are again angles and are the generalization of 
the definition of $\Omega_2$ in the 
3-dimensional space to $\hat\Omega_{A-2}$ in the $(A-1)$-dimensional space. Summarizing, the {\it hyper}-spherical coordinates 
include one {\it hyper}-radius  $\rho$ and $(3A-4)$ {\it hyper}-angles. 
We will denote all {\it hyper}-angles by $\hat\Omega_A$. So any function of the Jacobi coordinates 
$f(\vec\eta_1,\vec\eta_2,...\vec\eta_{A-1})$, when expressed in {\it hyper}-spherical coordinates becomes 
$f(\rho,\hat\Omega_{A})$.

The nice feature of these coordinates is that the kinetic energy operator
of a many particle system takes a form in perfect analogy to the 2-particle 3-dimensional case: 
a $\rho$-dependent Laplacian and a {\it hyper}-centrifugal barrier, with a  {\it hyper}-angular 
momentum operator $\hat{\bf K}$ that depends on all the {\it hyper}-angles. Therefore the translational invariant
Hamiltonian for an $A$-particle system reads.
\begin{equation}\label{H^[A]}
  H^{[A]} = - \frac{1}{2m}\Delta_{\rho}+ \frac{1}{2 m} \frac{\hat{\bf K}_A^2}{\rho^2}+ V^{[A]}(\rho,\hat\Omega_A)\,,
\end{equation}
 
The  hyperspherical harmonics $\cal Y_{[\it K]}$ are eigenfunctions of $\hat{\bf K}$ with eigenvalues
$K(K + 3A-5)$. They constitute a useful basis where one can expand the $A$-particle  wave function. 
It is as well useful to expand the hyperradial part in Laguerre polynomials $ L^\alpha_n(\rho)$
so that one has 
\begin{equation}
 \Psi(\rho,\hat\Omega_A) = \sum_{[K] n} C_{n [K]} L^\alpha_n(\rho) {\cal Y}_{[K]}(\hat\Omega_A) \,,
\end{equation}
Of course the wave function must be complemented by the spin-isospin parts. The whole function must be antisymmetric.
Therefore one needs to classify the  hyperspherical harmonics according to the irreducible representations of the 
symmetry group of $A$ particles. This is a non trivial task, that, however, has been solved in \cite{Nir9798}.

\subsection{Determination of an effective interaction for local potentials}  

In order to construct an EI we use the Lee-Suzuki similarity transformation~\cite{Suzuki}. This requires 
the definition of an $A$-particle model space $P$ ($Q$ denotes the rest of the whole Hilbert space, $P+Q=I$).
For us $P$ will be spanned by all the $A$-body HH with $K\leq K_{max}$. In order to get the {\it total} EI 
the Lee-Suzuki method gives the recipe to get the similarity transformation that defines $H^{[A]eff}$, 
and therefore $V^{[A]eff}$. 
However, as already said above, we do not search for the {\it total} EI,  but for a {\it partial} EI 
that is tailored for our HH model space and constrained to coincide with the bare one, only when enlarging the model space. 
This is naturally achieved  in the following way:
\begin{itemize}
 \item consider  $\rho$ as a parameter;
 \item focus on a part of $H^{[A]}$, that we will call $H^{[2]}(\rho)$,
\begin{equation}\label{H^[2]}
  H^{[2]}(\rho) = \frac{1}{2 m} \frac{\hat{\bf K}_A^2}{\rho^2}+ V_{A,A-1}(\rho,\hat\Omega_{A,A-1})\,.
\end{equation}
The term $V_{A,A-1}$  is just any of the pair potential terms in the Hamiltonian. For simplicity we choose 
the pair defining the 
''simplest' Jacobi coordinate $\vec\eta_{A,A-1} = \sqrt{\frac{1}{2}}(\vec{r}_{A}-\vec r_{A-1})$,
since anyway we are working with antisymmetrized states. 
The eigenfunctions of  $H^{[2]}(\rho)$ will live in a space that is contained in the $A$-body  Hilbert space. There will be 
a $P_2$ contained in $P$  and a corresponding $Q_2$ contained in $Q$. Both of them are known since the eigenvalue problem for 
 $H^{[2]}(\rho)$ can be easily solved, being a sort of 2-body problem ``immersed in a medium'' ($H^{[2]}$ depends on $\rho$!);
 \item  apply the  Lee-Suzuki similarity transformation to get the information residing in $Q_2$ into $P_2$ and 
obtain the EI $V^{[2]\,eff}(\rho)$; 
 \item  use this EI in the $P$ space;
 \item increase the $P$ space to convergence.
\end{itemize}

\section{The effective interaction for non-local potentials}  
For non-local interactions one cannot proceed in the same way as described above. The reason is that
for non-local interactions $V_{A,A-1}$ is a function of $(\rho,\hat\Omega_{A,A-1};\rho',\hat\Omega'_{A,A-1})$. If, as before 
the variable $\rho$ is considered a parameter, for  fixed $\rho$ and $\rho'$ one has
$V_{A,A-1}(\rho,\hat\Omega_{A,A-1};\rho',\hat\Omega'_{A,A-1})
\neq V_{A,A-1}(\rho,\hat\Omega'_{A,A-1};\rho',\hat\Omega_{A,A-1})$. This means that the potential is not hermitian!

A solution to overcome this problem is to start again from the full $H^{[A]}$, which for a non-local potential 
is written as
\begin{eqnarray}\label{H^[A]new} \nonumber
 H^{[A]} &=& - \frac{1}{2m}\Delta_{\rho}+ \frac{1}{2 m}
            \frac{\hat{\bf K}_A^2}{\rho^2}+ V^{[A]}(\rho,\hat\Omega_A;\rho',\hat\Omega'_A)\\
         &=& - \frac{1}{2m}\Delta_{\rho}+ \frac{1}{2 m}
            \frac{\hat{\bf K}_A^2}{\rho^2}+ V_{12}(\rho,\hat\Omega_{12};\rho',\hat\Omega'_{12})+...
             V_{A,A-1}(\rho,\hat\Omega_{A,A-1};\rho',\hat\Omega'_{A,A-1})\,,
\end{eqnarray}
and  subtract and add a term that is local only in $\rho$, i.e. $V_{A,A-1}(\rho,\hat\Omega_{A,A-1};\rho,\hat\Omega'_{A,A-1})$.
\begin{eqnarray}\label{H^[A]newnew}\nonumber 
 H^{[A]} = &-& \frac{1}{2m}\Delta_{\rho}+ \frac{1}{2 m}
            \frac{\hat{\bf K}_A^2}{\rho^2}+ V^{[A]}(\rho,\hat\Omega_A;\rho',\hat\Omega'_A)\\
 \nonumber        = &-& \frac{1}{2m}\Delta_{\rho}+ \frac{1}{2 m}
            \frac{\hat{\bf K}_A^2}{\rho^2}+ V_{12}(\rho,\hat\Omega_{12};\rho',\hat\Omega'_{12})+...
             V_{A,A-1}(\rho,\hat\Omega_{A,A-1};\rho',\hat\Omega'_{A,A-1})-\\ \nonumber
           &-&  V_{12}(\rho,\hat\Omega_{12};\rho,\hat\Omega'_{12}) +
                V_{12}(\rho,\hat\Omega_{12};\rho,\hat\Omega'_{12}) - ... + ...\\
           &-&  V_{A,A-1}(\rho,\hat\Omega_{A,A-1};\rho,\hat\Omega'_{A,A-1}) +
                  V_{A,A-1}(\rho,\hat\Omega_{A,A-1};\rho,\hat\Omega'_{A,A-1})\,.
\end{eqnarray}
\begin{table}[h]
\caption{\label{tb:4HeN3LO}
Convergence of the HH expansion for the $^4$He ground-state energy (in MeV) and
root-mean-square radius (in fm) with the bare and the effective non-local Idaho N$^3$LO potential.
Results with other methods are also given for comparison. 
}\bigskip
\begin{center}
\begin{tabular}
{c | cc | cc} \hline \hline
              &  \multicolumn{2}{c}{Bare} 
              &  \multicolumn{2}{|c}{Effective}  \\ \hline
    $K_{\rm max}$ &  $< H >$  &  $ \sqrt{ < r^2>} $
              &  $< H >$  &  $ \sqrt{ < r^2>} $
              \\ \hline
  2 &  -3.507 & 1.935 & -17.773  &   1.620    \\
  4 & -13.356 & 1.523 & -22.188  &   1.533    \\
  6 & -20.135 & 1.446 & -24.228  &   1.496    \\
  8 & -23.721 & 1.451 & -25.445  &   1.498    \\
 10 & -24.617 & 1.470 & -25.363  &   1.506    \\
 12 & -25.115 & 1.491 & -25.439  &   1.515    \\
 14 & -25.259 & 1.501 & -25.398  &   1.516    \\
 16 & -25.310 & 1.509 & -25.390  &   1.518    \\
 18 & -25.359 & 1.513 & -25.385  &   1.518    \\
 20 & -25.370 & 1.515 & -25.381  &   1.518    \\  
\hline 
    & -25.37(2) & 1.515(4) & -25.38(1)  &   1.518(1)    \\  
\hline 
 HH \cite{Viviani07} & -25.38 & 1.516 & & \\
 FY \cite{Viviani07,Nogga02} & -25.37 &  -    & &\\
 NCSM \cite{Gazit09} & -25.39(1) & 1.515(2)   & &   \\
\hline \hline
\end{tabular}
\end{center}
\end{table}
From this expression one can single out a new  $H^{[2]}(\rho)$ formed by the hypercentrifugal term 
and the last term
\begin{equation}
H^{[2]}(\rho)  = \frac{1}{2 m} \frac{\hat{\bf K}_A^2}{\rho^2}+ 
V_{A,A-1}(\rho,\hat\Omega_{A,A-1};\rho,\hat\Omega'_{A,A-1})\,.
\end{equation}
In this {\it quasi} two-body Hamiltonian all non-local effects regarding the hyperspherical
coordinates are incorporated, while the  hyperradial part of the non-locality remains excluded.

Applying the Lee-Suzuki transformation to this new  $H^{[2]}(\rho)$ one obtains  $V^{[2]\,eff}_{[K][K']}(\rho)$. 
Then the effective non-local interaction is obtained by 
\begin{equation}\label{v2effrho}
  v^{[2] \rm eff}_{[K_A],[K'_A]}(\rho,\rho') = 
        v^{[2]}_{[K_A],[K'_A]}(\rho,\rho') 
      + \delta(\rho-\rho') \left( v^{[2] \rm eff}_{[K_A],[K'_A]}(\rho,\rho) -
         v^{[2]}_{[K_A],[K'_A]}(\rho,\rho) \right) \;.
\end{equation}
The procedure described here is perfectly in line with the point of view stated above: 
all what is left out in the definition of the {\it partial} EI is recovered by increasing the model space
$P$ up to convergence.
In particular, since Eq.~(\ref{v2effrho}) represents a non-local HH effective interaction derived 
from the diagonal hyper-radial 
matrix element of $H^{[2]}$ in the position representation, it can be generalized to an arbitrary 
hyperradial basis. The question is which of this choices
will lead to a better effective interaction, i.e. to a faster convergence of the HH expansion. 
This point has been studied in~\cite{EIHHnonl}. 

\section{A test of the EIHH for non-local interaction}
Here we present applications of the procedure described in the previous chapter to the case
of 4- and 6-nucleon systems.  In Table 1  we present results for ground-state energy and  
root-mean-square radius of $^4$He, obtained with the bare non-local Idaho N3LO potential~\cite{N3LO}
and with the corresponding effective interaction, using the hyperradial basis that leads 
to the best convergence~\cite{EIHHnonl}.   (In order to work with this force we
use a representation  on a HO basis)

One sees that Kmax = 8 for ground-state energy  and Kmax = 12 
for the radius are already sufficient to obtain a convergence accuracy of less than 1\%.

For $A=6$ nuclei the situation is illustrated in Fig. 1, for the
ground-state energies of $^6$He and $^6$Li with the bare JISP16 nuclear force~\cite{JISP16} and with the
corresponding non-local EI. One notices that while it is not possible to obtain
converged HH results for the bare interaction, the EI convergence is reached with a rather
small effective interaction model space. 
\begin{figure}[h]
\begin{center}
\includegraphics[width=28pc]{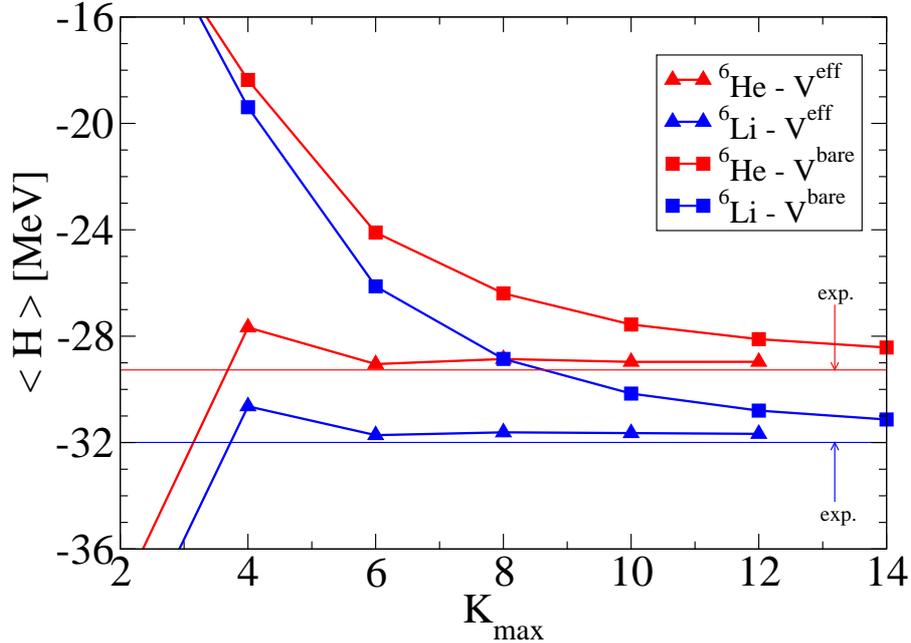}
\caption{\label{label} The ground-state energies of $^6$He and $^6$Li with the bare 
JISP16 nuclear force~\cite{JISP16} and with the corresponding non-local effective interaction.}
\end{center}
\end{figure}


\section{Perspectives for reaction cross sections}
Calculating a reaction cross section ab initio is considered a much more difficult task
than calculating binding energies or bound state observables in general. The reason is that in most cases 
reactions involve scattering states. The many-body
scattering problem may lack a viable solution already for a very small number
of constituents in the system. 

The Lorentz Integral Transform method~\cite{ELO94,ELOB07} is able to overcome this longstanding stumbling block.
In fact this approach reduces the scattering problem to a bound state-like problem.
The essence of the method lays in finding the solution of  a Schr\"odinger-like equation with a 
source $S$ that depends on the reaction one is study. In practice one has to solve 
\begin{equation}\label{LIT}
 \left(\hat{H}-E_0-\sigma\right)|\tilde{\Psi}\rangle=S\,,  
 \end{equation}
for many values of the real part $\sigma_R$ of the parameter $\sigma=\sigma_R+i\sigma_I$, and for a fixed value of its 
imaginary part $\sigma_I$, rigorously different from zero. It is just this last condition that ensures that the solution 
of Eq.~(\ref{LIT}) has   bound-state boundary conditions.  Therefore it can be found by any method suitable for bound states,
like that described in this contribution.

The results found in the application of the present version of the EIHH method  opens 
up a concrete  possibility to study structure and reaction of a few/many-particle system within the same framework,
also using potentials that are non-local, like the recent effective field theory potentials. This allows a critical 
review of different potential models, whose performances can be judged on a much larger set of observables.

\end{document}